\documentclass[aps,prl,twocolumn,showpacs]{revtex4-1}
\usepackage{graphicx}
\usepackage{dcolumn}
\usepackage{bm}
\usepackage{epstopdf}
\usepackage{amstext} 
\usepackage{fixmath} 

\begin{document}

\newcommand{\WT}[1]{\textbf{\textcolor{blue}{(#1 -- WT)}}}
\newcommand{\smco}{SmCo$_5$}
\newcommand{\w}{$\omega$}
\newcommand{\m}{$\mu$}

\title{Constraints on axionlike dark matter with masses down to $10^{-23}$~eV/c$^2$}

\author{W.\,A.~Terrano}
\altaffiliation{Present Address: Department of Physics, Princeton University, Princeton NJ 08550 USA}
\email{wterrano@princeton.edu}
\author{E.\,G.~Adelberger}
\author{C.\,A.~Hagedorn}
\author{B.\,R.~Heckel}
\affiliation{Center for Experimental Nuclear Physics and Astrophysics, Box 354290,
University of
Washington, Seattle, Washington 98195-4290} 

\begin{abstract}
We analyzed an 6.7-year span of data from a rotating torsion-pendulum containing $\approx 10^{23}$ polarized electrons to search for the ``wind'' arising from ultralight, axionlike dark matter with masses between $10^{-23}$ and $10^{-18}$~eV/c$^2$. Over much of this range we set a 95\% confidence limit $F_{\rm a}/C_{\rm e} > 2 \times 10^{15}$~eV on the axionlike decay constant.
\end{abstract}
\maketitle

A wide variety of astrophysical observations provide compelling indirect evidence for the existence of cold dark matter\cite{be:10}.
Direct efforts to detect this dark matter typically assume that it consists of heavy, fermionic, supersymmetry-inspired particles called WIMPS, or else low-mass bosons called axions that would solve the strong CP problem.  
Despite much effort,
large-scale detectors have not found evidence for supersymmetry\cite{supersymmetry} nor for dark-matter WIMPS\cite{wimps}. This has focused attention on low-mass bosonic dark matter, where sensitive instruments are now probing the expected coupling-strength and mass of the Peccei-Quinn axion\cite{admx}. 

Recent work\cite{gr:13, gr:18} has emphasized that bosons with masses anywhere between 10$^{-22}$ and 100~eV/c$^2$ 
could have been produced in the early Universe with the properties required of the cold dark matter.
If the bosons have masses below $1$~eV/c$^2$ and comprise a significant fraction of the observed dark matter density $\rho_{\rm DM}$, their number density must be so high that they behave as coherent waves rather than as particles. If the bosons are bound in our galaxy, they must be highly non-relativistic ($v_{\rm a}/c\approx 10^{-3}$), and the bosonic waves would have a frequency 
\begin{equation}
f_{\rm a}=E_{\rm a}/h=\frac{m_a c^2}{h} [1+(v_{\rm a}/c)^2/2],
\end{equation} 
corresponding to a central frequency  $f_{\rm a}=E_{\rm a}/h=m_{\rm a} c^2/h$ with a fractional spread $\delta f_{\rm a}/f_{\rm a}=(v_{\rm a}/c)^2/2 \approx 10^{-6}$, and a de Broglie wavelength\begin{equation}
\lambda_{\rm a}=h/(m_{\rm a} v_{\rm a})~.
\end{equation} 

Astrophysical observations may favor the very longest wavelength dark-matter candidates\cite{hu:00, hu:17, vi:18}:
conventional cold-dark-matter simulations predict density cusps at the centers of galaxies and a substantial abundance of low-mass dwarf galaxies, both of which disagree with observations\cite{CDM}.
An ultralight boson (UB) with $m_{\rm a} \approx 10^{-22}$~eV/c$^2$ would have $\lambda_{\rm a}\approx 1.2\times 10^{19}$~m or 400 parsecs. In this case, the uncertainty principle (which states that the UB distribution cannot be localized to better than $\lambda_{\rm a}$) could solve both of the above-mentioned problems with cold-dark-matter simulations, making this tiny mass an important target for experimental work.

Laboratory probes of axionic or axionlike dark-matter fall into three broad classes: coupling to highly sensitive electromagnetic 
circuits\cite{admx}, oscillating atomic\cite{ro:15} or neutron\cite{ab:17} electric-dipole or parity-violating moments, and ``wind''\cite{ab:17} effects. The ``wind'' from the laboratory's motion with respect to the axion wave acts on an electron like an oscillating ``magnetic field''\cite{gr:13, gr:18, st:14, al:18} with
\begin{equation}
 H_{\rm eff} = C_{\rm e}\frac{\tilde{a}_0}{2 F_{\rm a}}  \sin(2\pi f_{\rm a}  t+ \phi_{\rm a}) \frac{ (\mathbold{v_{\rm a} \cdot \sigma_{\rm e}})}{c}~,
\label{Heff}
\end{equation}
where the dimensionless factor $C_{\rm e}$ characterizes the axion coupling to electrons and $\mathbold{\sigma_{\rm e}}$ is the orientation of the electron spin. $F_{\rm a}$, $f_{\rm a}$, and $\phi_{\rm a}$ are the symmetry-breaking scale, oscillation frequency, and phase of the axionlike wave, respectively. If the local energy-density of dark matter (0.45 GeV/cm$^3$\cite{admx}) consists entirely of axionlike UBs, then $\tilde{a}_0=\sqrt{2 \rho_{\rm DM}(\hbar c)^3}\approx 2.6 \times 10^{-3}$~eV$^2$, a value we assume throughout this work.

Although the UBs and the solar system are gravitationally bound in the galaxy and originally had similar velocities this will not be the case today. The traditional isothermal isotropic equilibrium halo model\cite{os:74} predicts that dissipation and angular momentum conservation  (the ``ballerina effect") combined to give the present-day solar system a circular velocity $\mathbold{v}_{\odot}$ 
which is an order of magnitude larger than that of the dark matter. In this equilibrium model $\mathbold{v}_{\rm a}\approx -\mathbold{v}_{\odot}$. However, this simple model ignores the possibility that recent mergers between the Milky Way and its satellite galaxies have not  fully equilibrated, leaving debris streams with large velocities. Recent analyses of Gaia-2 data\cite{ev:18,fa:18,ne:18} suggest that 10-50\% of the dark matter in our galaxy is in such streams. If the solar system is currently in a debris stream, $\mathbold{v}_{\rm a}$ could point in any direction with a magnitude less than the local escape velocity.

\begin{figure}[t]
\includegraphics[width=.30\textwidth]{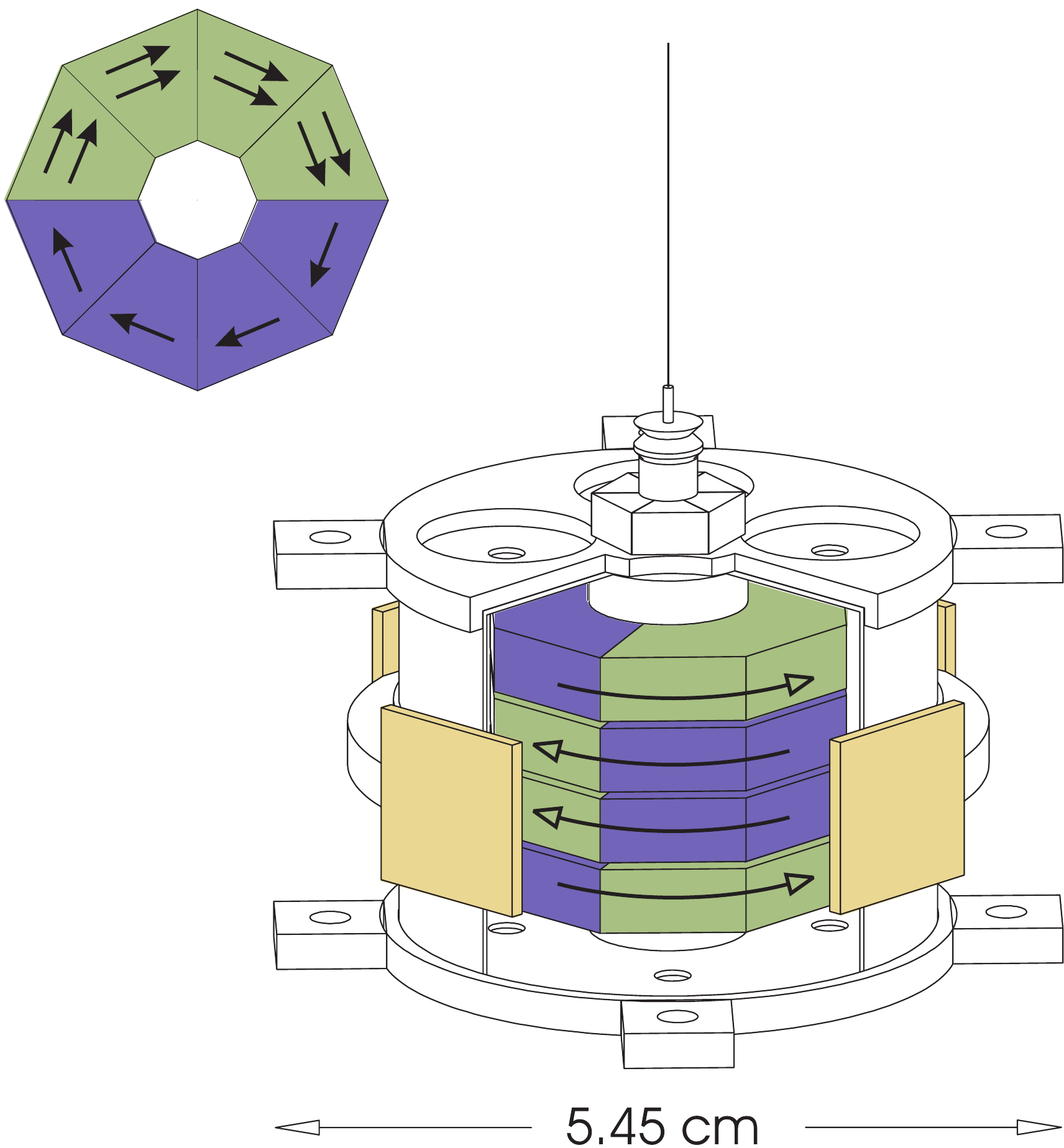}
\caption{(color online) Spin pendulum. The light green and darker blue volumes are Alnico and Sm$\,$Co$_5$, respectively. Upper left: top view of a single magnetized ``puck''; its spin moment points to the right. Lower right: the assembled pendulum with its magnetic shield shown cut-away to reveal the 4 pucks inside. Two of the 4 mirrors (light gold) used to monitor the pendulum twist are prominent. Arrows with filled heads show the relative densities and directions of the electron spins, open-headed arrows show the directions of ${\bm B}$. The pucks are arranged so that the spin dipole is centered on the pendulum, and the different materials have a vanishing composition-dipole moment. The 8 tabs on the shield held small screws that were used to tune out the pendulum's residual $q_{21}$ gravitational moment. The 103 g pendulum had a rotational inertia of 169 g$\,$cm$^2$. }
\label{pend}
\end{figure}

Motivated by these considerations, we searched for the frequency-dependent signatures of an axion wind in E\"{o}t-Wash rotating torsion-balance data previously taken with a pendulum containing $N_e\approx 9.8 \times 10^{22}$ polarized electrons.  This remarkable 
pendulum\cite{he:08} (shown in Fig.~\ref{pend})  was formed from closed magnetic circuits containing two different permanent magnetic materials with high (Alnico) and low (Sm$\,$Co$_5$) spin densities at the same internal magnetizations. The resulting device had a negligible external magnetic field but carried both net spin $S$ and orbital angular momentum $L$, with a total angular momentum $J=-S$. The net spin dipole was calibrated using the Coriolis force from the earth's rotation on the electron-spin  ``quantum gyroscope''. The device was sufficiently sensitive to yield an upper limit of $\approx 2 \times 10^{-22}$ eV on the energy required to invert an electron spin about a direction fixed in inertial space; this is equal to the electrostatic energy of 2 electrons separated by 48 astronomical units.  However that analysis, which searched for a preferred frame, would have averaged away the time-varying signal of Eq. \ref{Heff}.

%

Here we analyze a larger data set (taken for Refs.~\cite{he:08,te:11,he:13}) that spanned 2438 days in 241 subsets. During each subset the experimental conditions (for example the turntable rotation rate, the angle of the spin dipole in the turntable frame\cite{he:08,te:11,he:13} and the positions of external sources\cite{te:11,he:13}) remained constant. The subsets had durations between 0.8 and 10 days. 
Our data consisted of 15588 individual measurements. Each measurement typically contained exactly 2 full turntable revolutions and lasted for $\sim 2800$\,s, a duration  long compared to the pendulum's free-oscillation period of $\approx 200$\,s. Following Ref.~\cite{he:08}, we assumed that during an individual measurement the pendulum's energy as a function of turntable angle $\phi_{\rm tt}$ was
\begin{equation}
E(\phi_{\rm tt})=-N_e \mathbold{\sigma \cdot \beta}=-N_e \beta\cos\phi
\end{equation}
where $N_e$ is the number of polarized electrons in the pendulum, $\mathbold{\sigma}$ is the direction of the spin dipole, 
$\mathbold{\beta}$ is a vector assumed to be approximately fixed in the lab during an individual measurement, and $\phi=\phi_{\rm tt}-\phi_0$ was the instantaneous angle between the rotating spin dipole and $\mathbold{\beta}$. Therefore the pendulum experienced a torque
\begin{equation}
T(\phi_{\rm tt})=-\partial E/\partial \phi=N_e(\beta_{\rm W} \cos \phi_{\rm tt} -\beta_{\rm N} \sin \phi_{\rm tt})
\end{equation}
that was inferred by correcting the measured pendulum twist angle in the rotating frame for pendulum inertia plus electronic and digital
time constants.
Each measurement yielded independent determinations of $\beta_{\rm N}$ and $\beta_{\rm W}$, where N and W are local North and West directions. Measurement uncertainties in each data subset were deduced from the scatter of the points in that subset. 
%
\begin{figure}
\includegraphics[width=.45\textwidth]{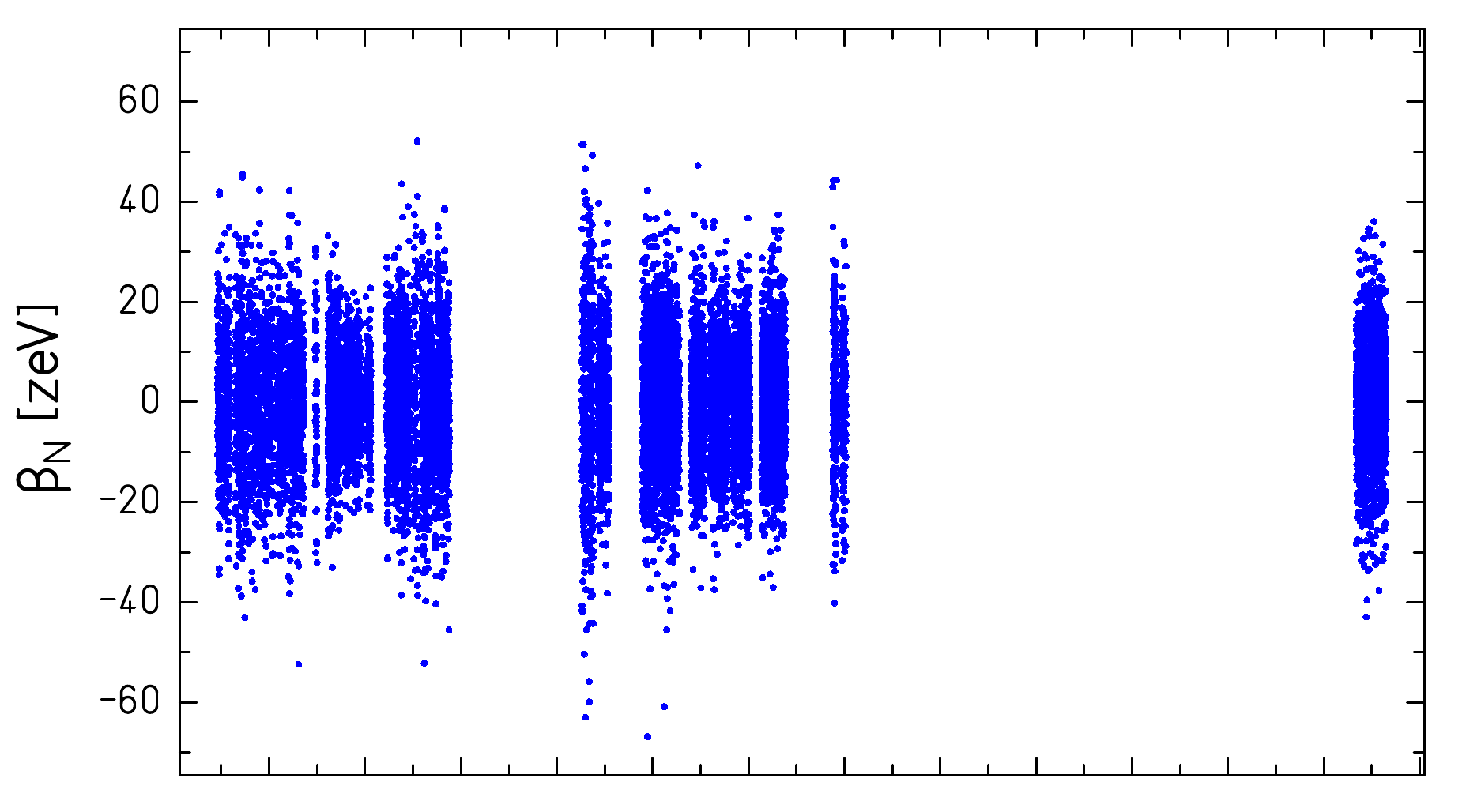}  	
\includegraphics[width=.45\textwidth]{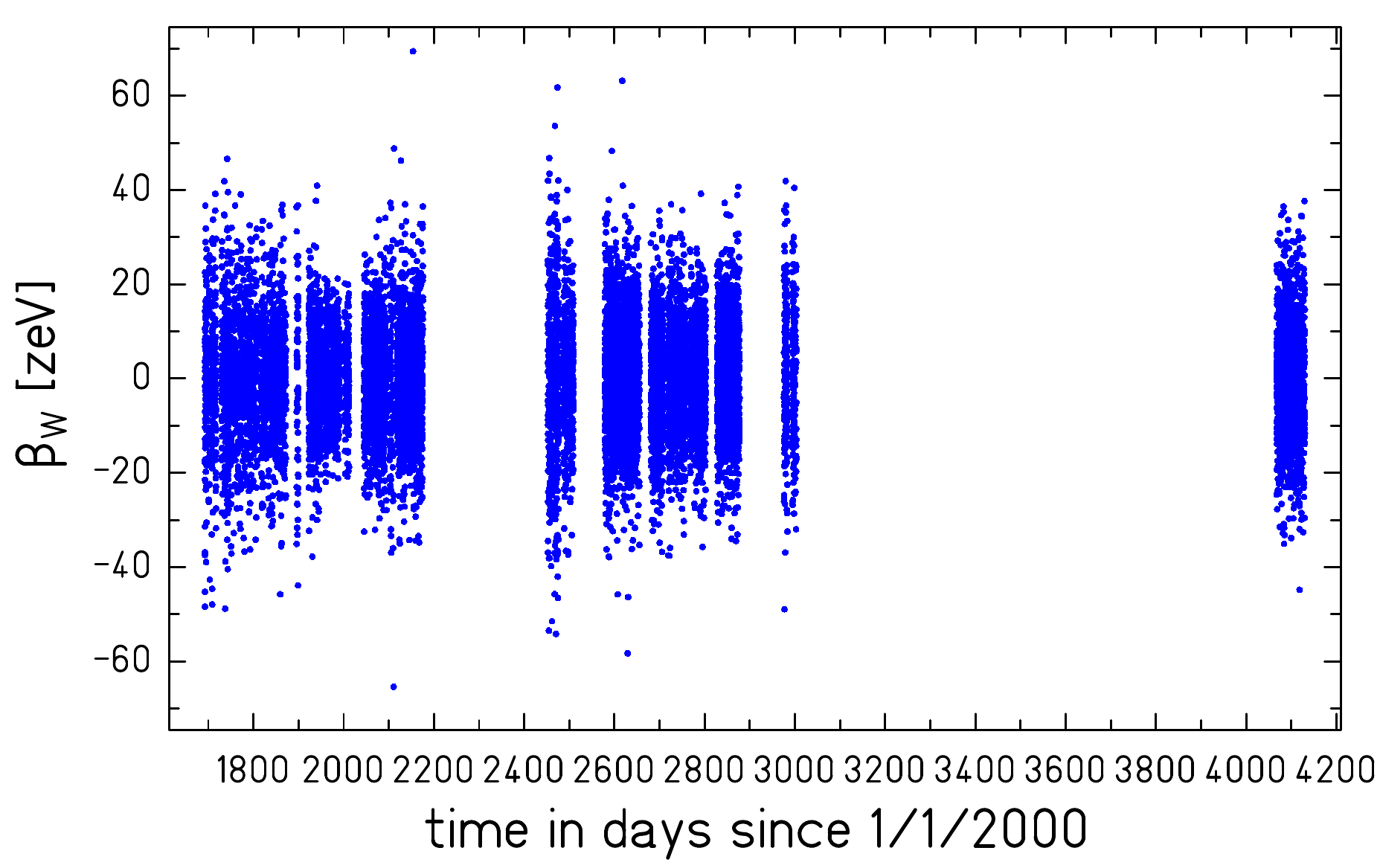}  	
\caption{$\beta_{\rm N}$ and $\beta_{\rm W}$ values from all 241 subsets. The different subsets have varying noise levels which are reflected in the assigned uncertainties (not shown).}
\label{betaW}
\end{figure}
We suppressed lab-fixed signals (arising from the Coriolis force as well as from many other less interesting effects) by setting to zero the average $\beta_{\rm N}$ and $\beta_{\rm W}$ values in each of the 241 subsets. Figure~\ref{betaW} displays the $\beta_{\rm N}$ and $\beta_{\rm W}$ values of our measurements.  

We searched
our $\beta_{\rm N}$ and $\beta_{\rm W}$ values for signals from axions with $\mathbold{v}_{\rm a}$ in an arbitrary direction in the equatorial (X,Y) plane. (Signals for $\mathbold{v}_{\rm a}$ along Z were not considered here as they have no sidereal modulation and are more difficult to distinguish from mundane lab-fixed effects\cite{lab fixed}.) We made evenly-spaced scans over 67,200 values of $f_{\rm a}$. At each $f_{\rm a}$ we computed 8 basis states: 
$b^i_{\rm XN \cos}$, $b^i_{\rm XW \cos}$, $b^i_{\rm XN \sin}$, $b^i_{\rm XW \sin}$, plus corresponding states for $\mathbold{v}_{\rm a}$ along $\mathbold{\rm Y}$. Here, for example, $b^i_{\rm XNcos}=K^i_{\rm XN}\eta^i\cos{\omega t^i}$ where $K^i_{\rm XN}$ transforms transforms equatorial (X,Y) to local (N,W) coordinates and varies at the sidereal frequency; $\eta^i=\sin(\omega\tau^i/2)/(\omega \tau^i/2)$ accounts for the attenuation of $f_{\rm a}$ by the finite length of the measurements: $t^i$ and $\tau^i$ are the mean time and duration of the $i^{\rm th}$ measurement, and $\omega=2\pi f_{\rm a}$. We zeroed the average values of the 8 basis states in each of the 241 subsets as well. 
This procedure ensured that the effects of zeroing the mean values of the $\beta$ data subsets, the varying lengths and uncertainties of the subsets, as well as the gaps between the subsets were handled correctly.  
Our constraints for axion velocities along X were derived from a linear fit that yielded quadrature amplitudes  $a_{X\!\!\cos}$ and $a_{X\!\!\sin}$
\begin{eqnarray}
\beta_{\rm N}^{\: i}&=& a_{\rm X\!\!\cos} \: b_{\rm XNcos}^i + a_{\rm X\!\!\sin} \: b_{\rm XNsin}^i   \nonumber  \\
\beta_{\rm W}^{\: i}&=& a_{\rm X\!\!\cos} \: b_{\rm XWcos}^i + a_{\rm X\!\!\sin} \: b_{\rm XWsin}^i ~,
\label{eq: def}
\end{eqnarray}
with similar expressions for $a_{\rm Y\!\!\cos}$ and $a_{\rm Y\!\!\sin}$. 
%
\begin{figure}[t]
\includegraphics[width=.40\textwidth]{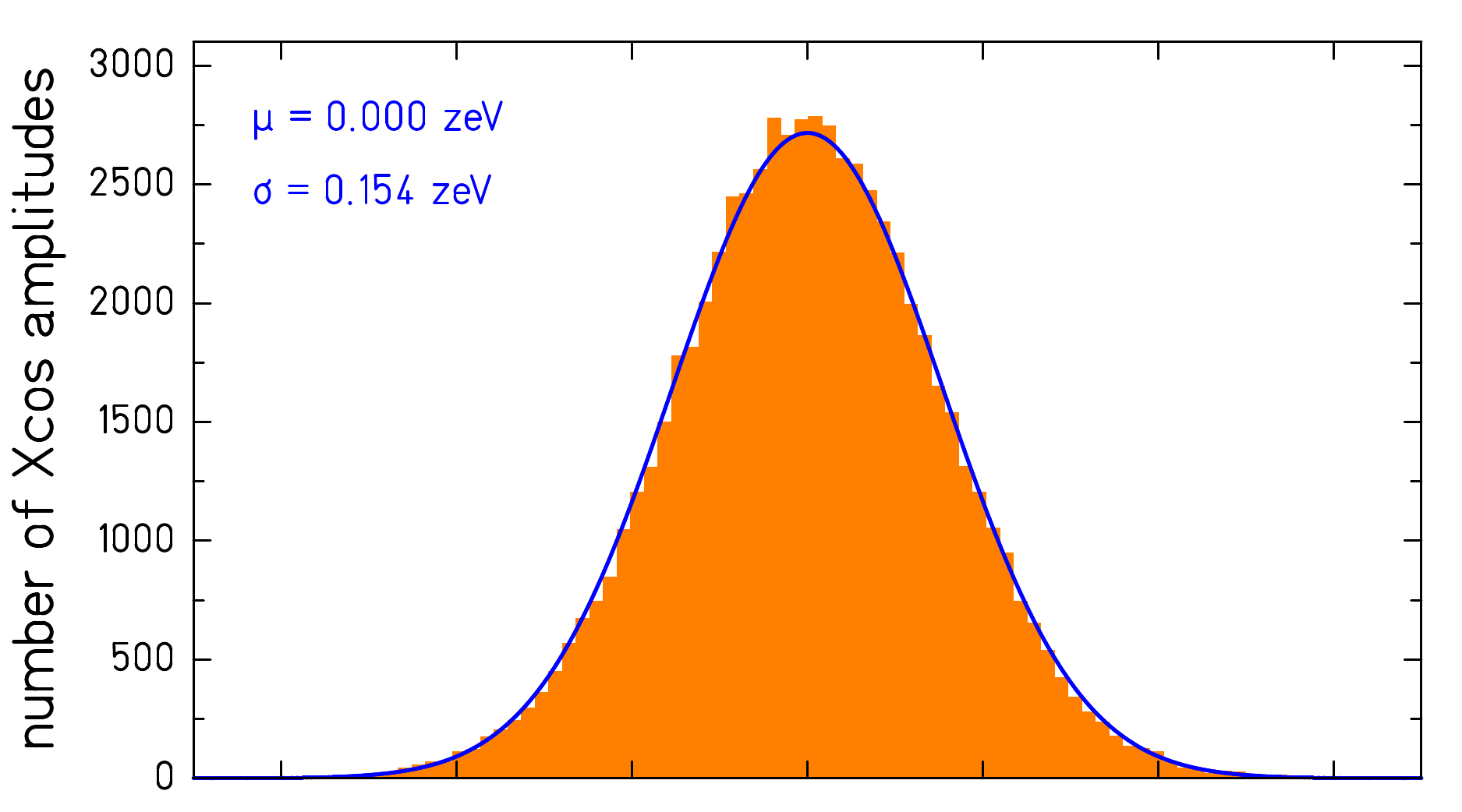} 	
\includegraphics[width=.40\textwidth]{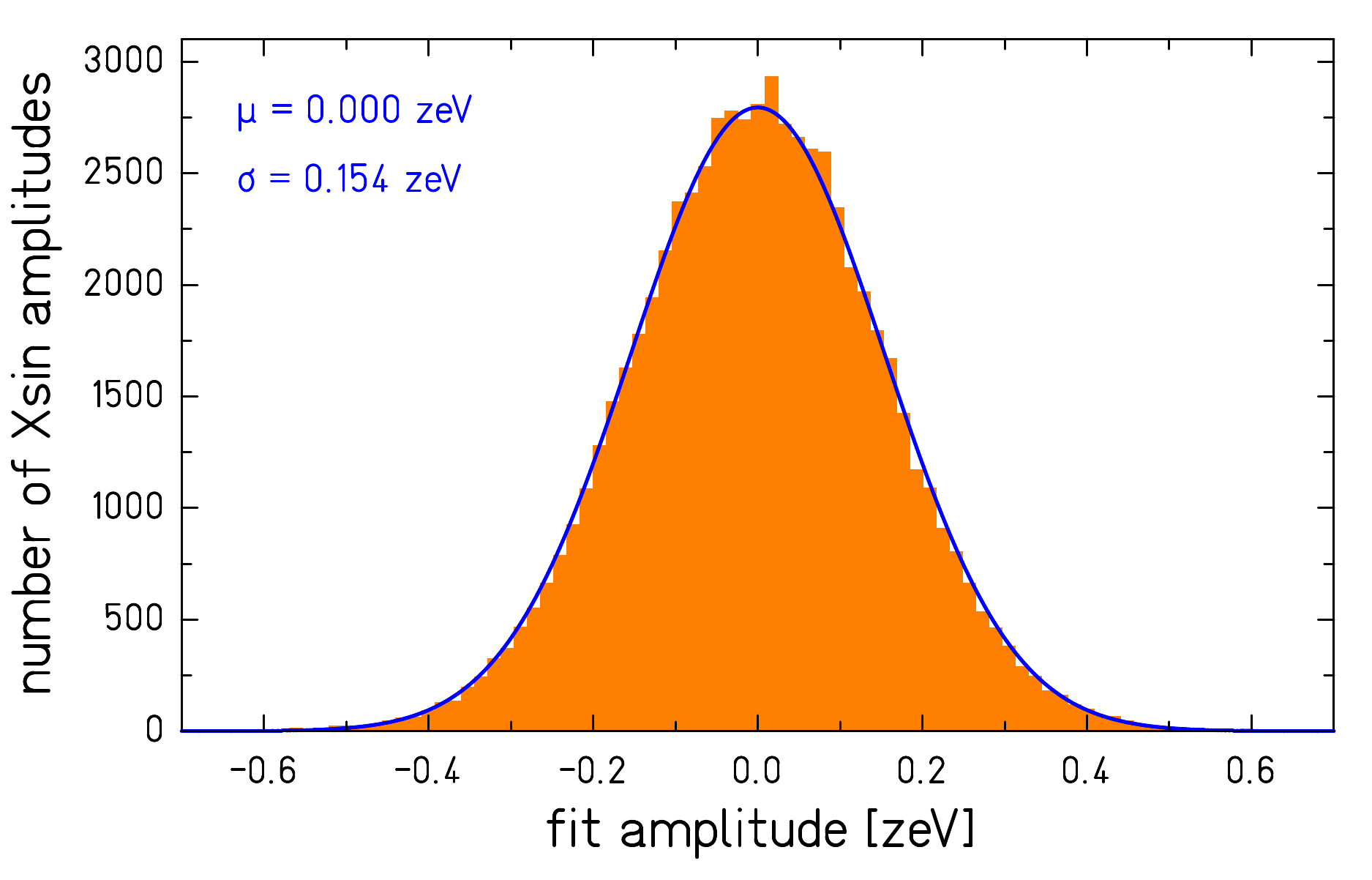} 	
\includegraphics[width=.40\textwidth]{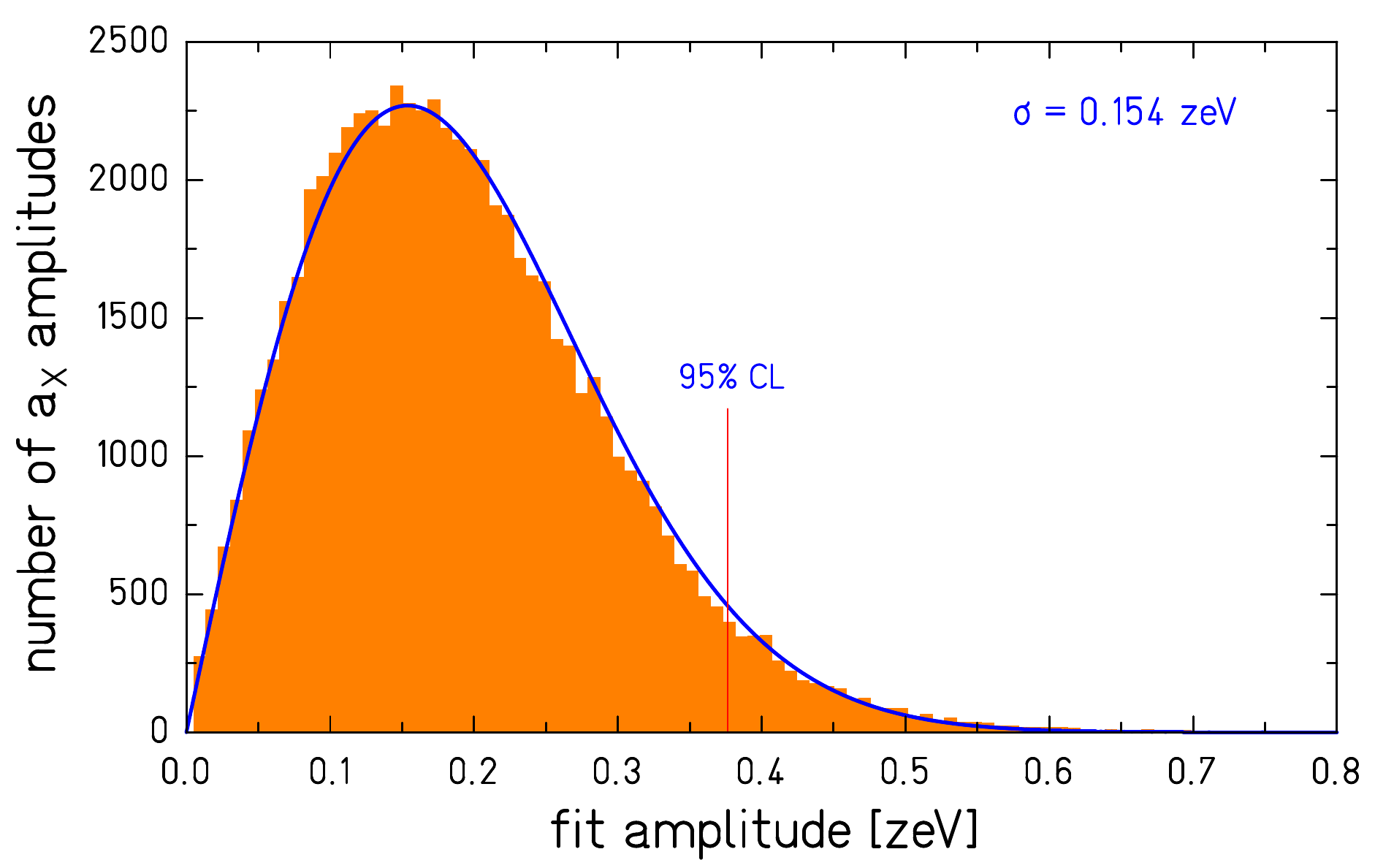} 	
\caption{(color online) Upper 2 panels: histograms of $a_{\rm X\!cos}$ and $a_{\rm X\!sin}$ coefficients for $10^{-8} \leq f_{\rm a} \leq 10^{-4}$~Hz. Results for $a_{\rm Y\!cos}$ and $a_{\rm Y\!sin}$ are very similar. All 4 quadrature amplitudes are characterized by zero-mean Gaussians with $\sigma=0.154$~zeV. Lower panel: histogram of corresponding $a_X$ amplitudes. The results follow the expected Rayleigh distribution, the 95\%-confidence upper-limit on
$a_{\rm X}$ (as well as on $a_{\rm Y}$) is 0.38~zeV. }
\label{xhist}
\end{figure}

We first analyzed the frequency range of our highest sensitivity ($1\times 10^{-8}\le f_{\rm a} \le 1\times 10^{-4}$ Hz); here the maximum $f_{\rm a}$ was small compared to the inverse durations of the measurements so that the basis states were not appreciably attenuated.  For each $f_{\rm a}$, we determined 4 quadrature amplitudes and their uncertainties. 
Results are shown in Figure~\ref{xhist}.
All 4 quadrature amplitudes are characterized by Gaussians with zero mean and $\sigma=0.154$~zeV. The distributions of the quadrature amplitudes divided by their uncertainties are also zero-mean Gaussians but with $\sigma=1.0$, We marginalised over the uninteresting phase $\phi_{\rm a}$ by computing radial amplitudes $a_{\rm X}=\sqrt{|a_{\rm X\!cos}|^2+|a_{\rm X\!sin}|^2}$ and  $a_{\rm Y}=\sqrt{|a_{\rm Y\!cos}|^2+|a_{\rm Y\!sin}|^2}$. As expected, $a_{\rm X}$ and $a_{\rm Y}$ are well modeled by the Rayleigh distribution whose only free parameter is the Gaussian $\sigma$.
%
%
\begin{figure}[t]
\includegraphics[width=.48\textwidth]{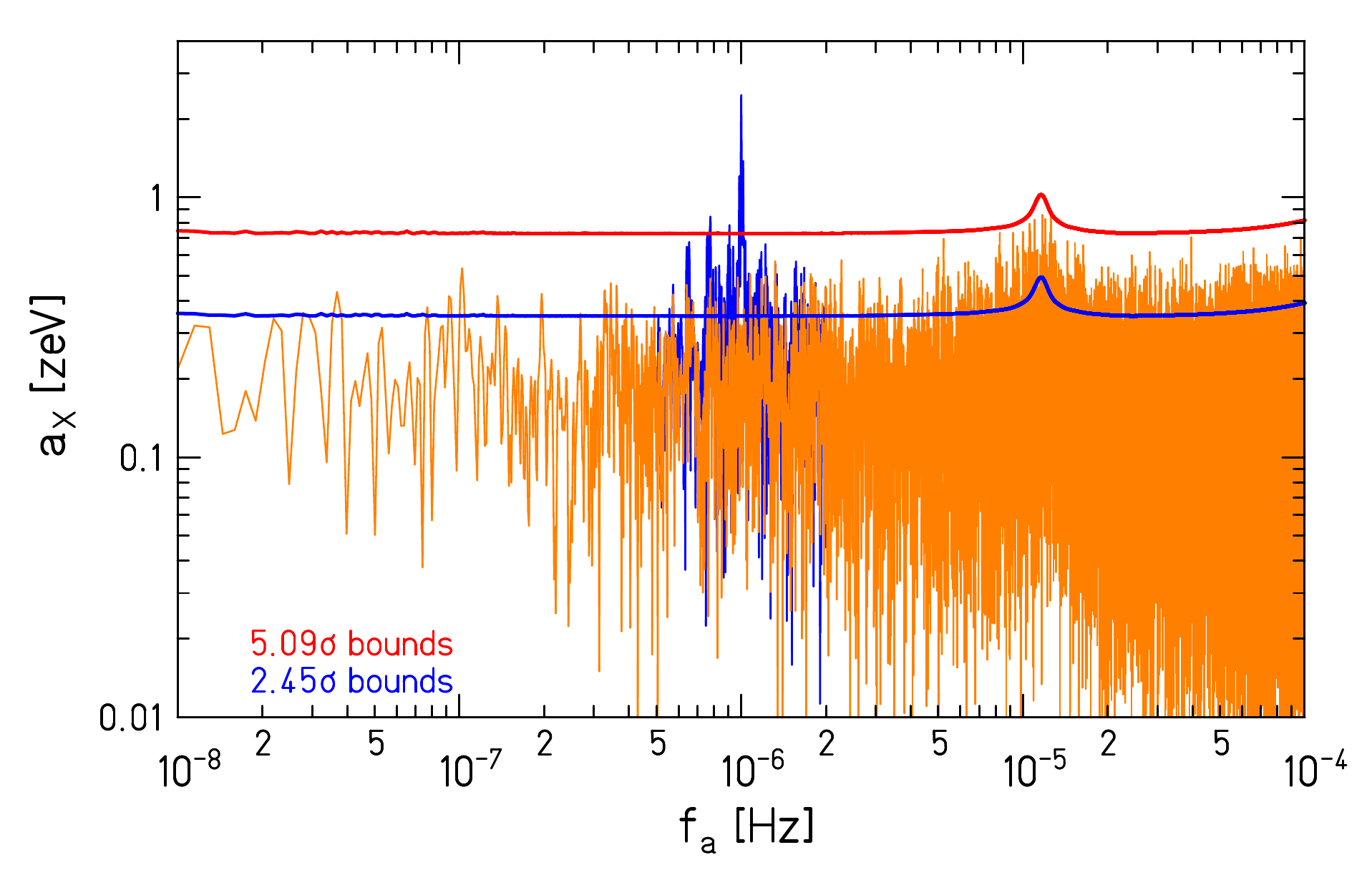}  	
\caption{(color online) 
Spectral distribution of $a_{\rm X}$. The lighter (orange) jagged line shows the fit amplitudes. The 2 smooth lines show the  sensitivities of the analysis. The lower (blue) smooth curve shows the 95\% C.L. Rayleigh uncertainties in the individual fit amplitudes while the upper red curve contains a trials penalty. Absent a signal, the Rayleigh distribution predicts that 5\% of the amplitudes will lie above the lower curve and that there is a 5\% probability that a single amplitude anywhere in our frequency range will lie above the upper curve; as expected no amplitude exceeds the upper curve.  
The darker (blue) jagged line shows the result of adding to our $\beta_N$ and $\beta_W$ data a synthetic 1$\mu$Hz $X_{\cos}$ signal with an amplitude of 2.5~zeV. The extracted amplitude, $2.498 \pm 0.144$~zeV, is resolved at $17\sigma$. Small satellite peaks in the synthetic signal arise from gaps in our time-series data.}
\label{xampdist}
\end{figure}

Figure~\ref{xampdist} shows the spectral distribution of $a_{\rm X}$. The small bump centered at $f_{\rm a}\approx 11.6~\mu$Hz occurs when $f_{\rm a}$ and the sidereal frequency coincide. In that case our procedure of zeroing the average values of the basis states reduced their mean magnitudes by a factor $\sqrt{2}$. This automatically increased the fitted amplitudes (and their uncertainties) by the same factor. A careful examination revealed that the central values in the bump region exceed the expected increase by an additional 40\%, which accounts for the observation that 5.2\%, rather than 5.0\%, of the points in Fig.~\ref{xampdist} lie above the lower blue curve.  We checked that the excess on the  bump was not an artifact of our analysis using a simulation where we kept the actual uncertainties and times of our $\beta$ data, but replaced the central values of the 
$\beta$s with values normally-distributed around 0 by the actual uncertainties. The simulated data showed no additional excess, and we conclude that the excess arose from a subtle systematic with a characteristic period of about 1 day. Binning the data as function of time of data did reveal a systematic effect: the scatter of the points in night-time data was less than that of the day-time data, but correcting for this had no significant effect on the results.
%

%
\begin{figure}[!t]
\includegraphics[width=.45\textwidth]{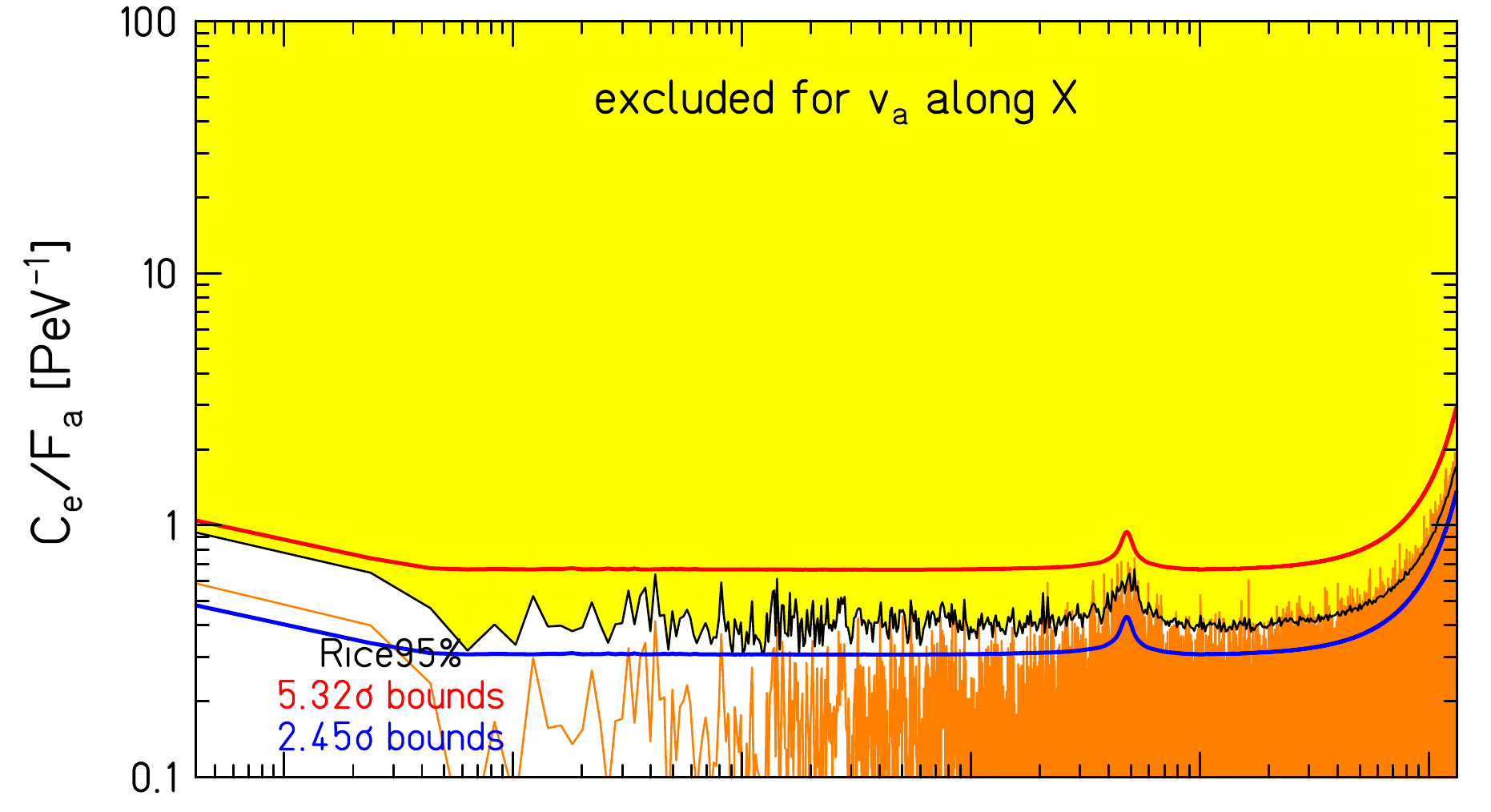} 	
\includegraphics[width=.45\textwidth]{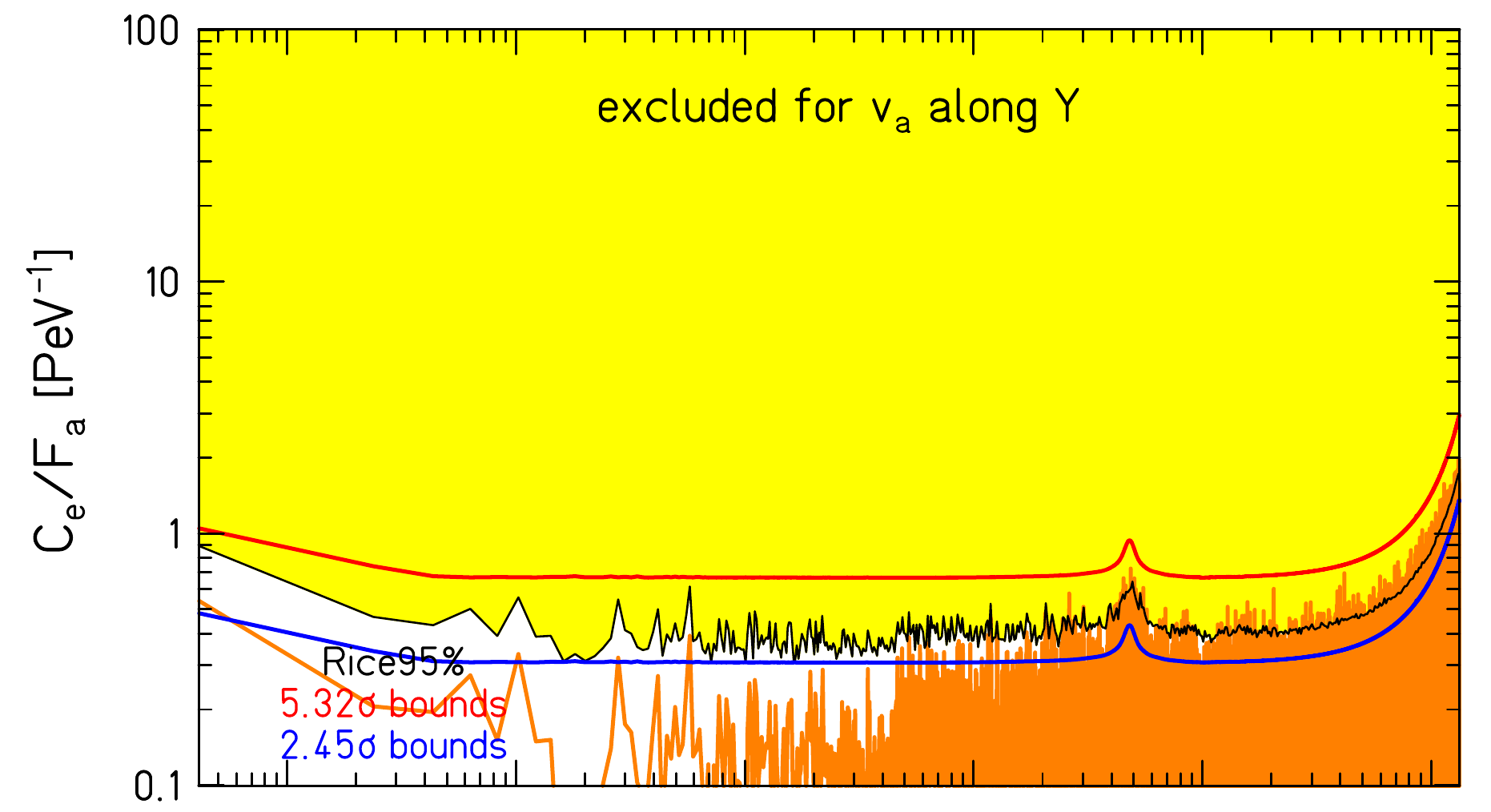} 	
\includegraphics[width=.45\textwidth]{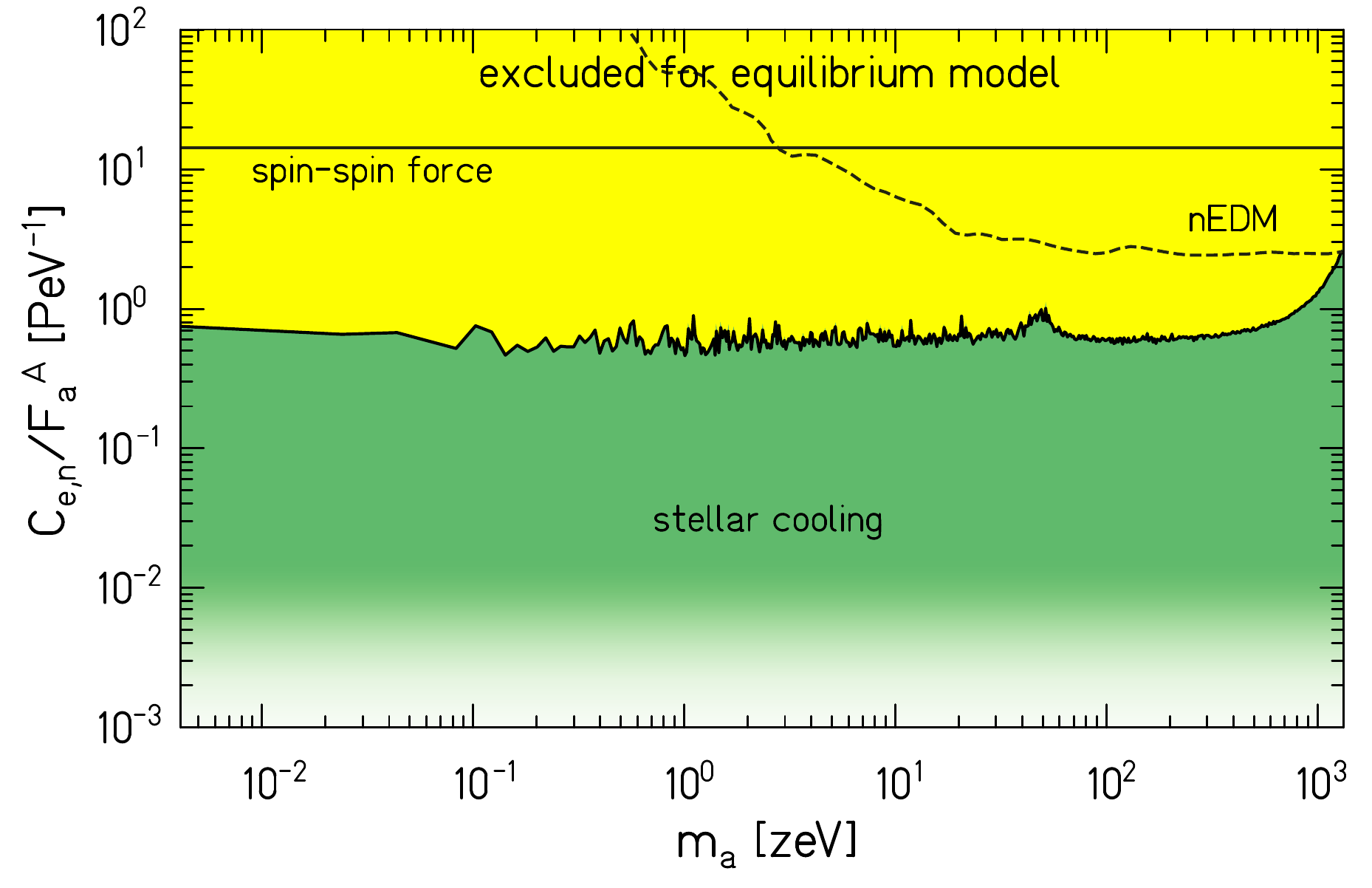}   	
\caption{(color online) 95\%-confidence constraints on $C_{\rm e}/F_{\rm a}$. The color coding of the lines is identical to that in Fig.~\ref{xampdist}.
The structure near $m_a$ = 47~zeV is the sidereal bump and the smooth increase at the highest masses reflects the signal attenuation due to finite measurement durations. The jagged black line shows the 95\% C.L. upper limit as a function of $m_{\rm a}$ computed from the Rice distribution\cite{rice} (the generalization of the conventional 
mean+2$\sigma$ appropriate for quadrature amplitudes). For clarity, we averaged Rice95 points with $m_{\rm a}> 2$ zeV. See Supplemental Material\cite{URL} for the unaveraged values. Absent a signal, the 
Rayleigh distribution predicts that 5\% of the unaveraged amplitudes (orange) should lie above the $2.45\sigma$ line (in fact for the top, middle and bottom panels 4.8\%, 5.2\% and 5.2\% do); and that there is only a 5\% chance that a single unaveraged amplitude anywhere the distribution should lie above the $5.32\sigma$ line (in fact none do). The dashed line shows $C_{\rm n}/F_{\rm a}$ constraints from an axion-wind analysis\cite{ab:17} of neutron electric-dipole data. The black horizontal line shoes the constraint from a spin-spin force experiment\cite{te:15}. The shaded green region is disfavored by stellar cooling\cite{WDcooling}.}  
\label{95CL}
\end{figure}
Satisfied that the statistical properties of $a_{\rm X}$ and $a_{\rm Y}$ were well described by the Rayleigh distribution, we widened our scan to cover frequencies between $10^{-9}$ Hz and $3.2\times 10^{-4}$ Hz. Now the frequency interval in our 67,200 point scan nearly equaled the inverse of the 2438-day span our our data and the high-frequency signals were appreciably attenuated by the finite durations of the measurements.
Our $C_{\rm e}/F_{\rm a}$ constraints from this scan on axion velocities lying in an arbitrary direction in the XY plane are shown in Fig.~\ref{95CL}.  We assume that $|\mathbold{v}_{\rm a}|=240$km/s $\approx|\mathbold{v}_{\odot}|$\cite{bo:12}, this is approximately the virial velocity and roughly half the local escape velocity\cite{pi:14}. We neglected the Earth's orbital velocity as it is an order of magnitude smaller than $|\mathbold{v}_{\odot}|$. 
Our constraints based on the simple equilibrium halo model, where $\mathbold{v}_{\rm a}$ is the XY component of 
$\mathbold{v}_{\odot}$, are shown in Fig.~\ref{95CL} as well.

This analysis probes axionlike dark-matter coupled to electrons over 5 of the 22 decades of possible masses for coherent axionlike dark-matter\cite{gr:13}. Were it of sufficient interest, one could re-fit the raw pendulum-twist data with combined turntable, dark-matter and sidereal basis-functions,  extending constraints up to the pendulum's free-oscillation frequency of $\approx 5$~mHz and  cover an additional decade of mass range.

Our results exclude ultralight galactic-halo axions with  $F_{\rm a}/C_{\rm e} \lesssim$ 2 PeV, significantly above the axion scales $F_{\rm a}/C_{\rm e} \lesssim$70 TeV  probed by experiments that rely on both sourcing and detecting axions, such as fifth-force searches\cite{he:13,te:15} and light-shining-through-walls tests\cite{ba:15}.  Axions in this parameter space would require a mechanism that suppresses their production inside stars, such as a chameleonic self-interaction\cite{ja:06}, since 
astrophysical considerations disfavor axions with $F_{\rm a}/C_{\rm e} \lesssim 1$~EeV  as they would provide a channel for excessive cooling\cite{WDcooling}. 
It is interesting that an EeV-scale axion would slightly improve the consistency between cooling models and observations for some well-studied astrophysical systems
\cite{Corsico2016,Viaux2013}. 

We expect that by replacing the tungsten torsion fibers used in this work with ones made from fused silica, and employing advanced twist-angle readout and turntable control, the constraints reported here could be improved by up to a factor 100. An additional factor of 10 would be needed to probe beyond the coupling strengths disfavored by astrophysics. We thank J. Detwiler, V. Flambaum, L. Hui, M. Lisanti, Y. Stadnik and T. Quinn for helpful conversations. Ted Cook constructed the spin pendulum and Claire Cramer took some of our data. This work was supported in part by National Science Foundation Grants PHY-1305726 and PHY-1607391.


\begin{references}
\bibitem{lab fixed}Ref.~\cite{he:08} obtained a constraint along Z, albeit with lower precision, by taking data with 4 orientations of the spin pendulum with respect to the torsion fiber, however, this was not done for the data taken for Refs.~\cite{te:11,he:13}.
\end{references}
\end{document}